\documentstyle[11pt,aaspp4,flushrt]{article}  
\journalid{VOL}{JOURNAL DATE 1996}
\articleid{START PAGE}{END PAGE}
             \font\sevenrm=cmr7

          \font\sixrm=cmr6       
                     
\def\teq#1{$\, #1\,$}                           
{\catcode`\@=11                                                  
\gdef\SchlangeUnter#1#2{\lower2pt\vbox{\baselineskip 0pt\lineskip0pt    
\ialign{$\m@th#1\hfil##\hfil$\crcr#2\crcr\sim\crcr}}}}           
\def\gtrsim{\mathrel{\mathpalette\SchlangeUnter>}}               
\def\lesssim{\mathrel{\mathpalette\SchlangeUnter<}}

\def\erg{\varepsilon}
\def\eperp{\vphantom{(}\erg_{\perp}}
\def\lambar{\lambda\llap {--}}
\def\fsc{\alpha_{\hbox{\sevenrm f}}}                                
\def\dover#1#2{\hbox{${{\displaystyle#1 \vphantom{(} }\over{
   \displaystyle #2 \vphantom{(} }}$}}                
\def\split{\gamma\to\gamma\gamma}
\begin{document}
%
%
\newcommand{\vol}[2]{$\,$\rm #1\rm , #2.}                 
\newcommand{\figureout}[2]{  \begin{figure}  \epsfysize=12.5cm
   \hbox to\hsize{\hfill \hbox{\epsfbox{#1}} \hfill} 
   \vskip 0.0cm \caption{#2} \end{figure} \clearpage }
\newcommand{\tableout}[4]{\vskip 0.3truecm \centerline{\rm TABLE #1\rm}
   \vskip 0.2truecm\centerline{\rm #2\rm}   
   \vskip -0.3truecm  \begin{displaymath} #3 \end{displaymath} 
   \noindent \rm #4\rm\vskip 0.1truecm } 
%
%
\title{MAGNETIC PHOTON SPLITTING: COMPUTATIONS OF \\
       PROPER-TIME RATES AND SPECTRA}
   \author{Matthew G. Baring\altaffilmark{1} and Alice K. Harding}
   \affil{Laboratory for High Energy Astrophysics, Code 661, \\
      NASA Goddard Space Flight Center, Greenbelt, MD 20771, U.S.A.\\
      \it baring@lheavx.gsfc.nasa.gov, harding@twinkie.gsfc.nasa.gov\rm}
   \altaffiltext{1}{Compton Fellow, Universities Space Research Association}
   \authoraddr{Laboratory for High Energy Astrophysics, Code 661,
      NASA Goddard Space Flight Center, Greenbelt, MD 20771, U.S.A.}
%
%

\begin{abstract}  
The splitting of photons \teq{\gamma\to\gamma\gamma} in the presence of
an intense magnetic field has recently found astrophysical applications
in polar cap models of gamma-ray pulsars and in magnetar scenarios for
soft gamma repeaters.  Numerical computation of the
polarization-dependent rates of this third order QED process for
arbitrary field strengths and energies below pair creation threshold is
difficult: thus early analyses focused on analytic developments and
simpler asymptotic forms.  The recent astrophysical interest spurred
the use of the S-matrix approach by Mentzel, Berg and Wunner to
determine splitting rates.  In this paper, we present numerical
computations of a full proper-time expression for the rate of splitting
that was obtained by Stoneham, and is exact up to the pair creation
threshold.  While the numerical results derived here are in accord with
the earlier asymptotic forms due to Adler, our computed rates still
differ by as much as factors of 3 from the S-matrix re-evaluation of
Wilke and Wunner, reflecting the extreme difficulty of generating
accurate S-matrix numerics for fields below about \teq{4.4\times
10^{13}}Gauss.  We find that our proper-time rates appear very
accurate, and exceed Adler's asymptotic specializations significantly
only for photon energies just below pair threshold and for
supercritical fields, but always by less than a factor of around 2.6.
We also provide a useful analytic series expansion for the scattering
amplitude valid at low energies.
\end{abstract}  
\clearpage 
\section{INTRODUCTION}

The exotic quantum electrodynamical (QED) process of the splitting of a
photon into two (\teq{\gamma\to\gamma\gamma}) in the presence of a
strong magnetic field has recently been of interest in the study of
gamma-ray pulsars and soft gamma repeaters (SGRs).  Many
radio pulsars are known to have spin-down fields in excess of
\teq{10^{12}}Gauss, with about a dozen exceeding \teq{10^{13}}G,
including the \teq{\gamma}-ray pulsar PSR1509-58.  This source is
unusual among the seven known \teq{\gamma}-ray pulsars in that it has a
hard power-law spectrum that turns over sharply around 1 MeV (Ulmer et
al. 1993; Hermsen et al. 1997), whereas other pulsars (e.g. the Crab;
see Nolan et al.  1993) whose fields are nearer \teq{10^{12}}G have
emission extending beyond 1 GeV.  A question then naturally arises:
does the high field of PSR1509-58 cause its spectrum to differ so much
from that of the other \teq{\gamma}-ray pulsars?  It appears that
magnetic photon splitting, in concert with the more familiar process of
single-photon pair creation, may provide the answer (Harding, Baring
\& Gonthier 1996, 1997) in the context of polar cap models, inhibiting
emission above a few MeV for a reasonable range of cap sizes.  Photon
splitting may also be quite important in the most recently discovered
(Ramanamurthy et al. 1996) gamma-ray pulsar, PSR0656+14, which has a
spin-down field of \teq{9.3\times 10^{12}} Gauss, between that of the
Crab and PSR1509-58.

Also of interest to high energy astrophysicists are soft gamma
repeaters (SGRs), the transient sources that are observed to have
sporadic periods of high \teq{\gamma}-ray activity.  There are three
known repeating sources, all with sub-second durations and soft spectra
(\teq{kT\sim 30} keV: for SGR1900+14, see Kouvelioutou et al. 1993).
The identification of radio (SGR1806-20: Kulkarni and Frail, 1993) and
X-ray (e.g. Mar 5, 1979 repeater: Rothschild, et al. 1994) counterparts
to SGRs spawned much interest in these objects.  It has been
suggested that SGRs are {\it magnetars}, neutron stars with extremely
high fields, in excess of the quantum critical field strength
\teq{B_{\rm cr} =m_e^2c^3/ e\hbar=4.413 \times 10^{13}} Gauss (when the
cyclotron and electron rest mass energies are comparable); the
spin-down estimate for the Mar. 5, 1979 repeater is \teq{B\sim 6\times
10^{14}} Gauss (Duncan and Thompson, 1992), assuming its age is that of
N49, its associated supernova remnant.  Such fields far exceed those
found in most radio pulsars, and can perhaps be generated by
enhancement due to dynamo action (Duncan \& Thompson, 1992).  Baring
(1995) proposed the potential importance of photon splitting for
softening emission spectra in SGRs via a splitting cascade to reproduce
the observed emission spectra, yielding an estimate of \teq{B\gtrsim
2\times 10^{14}} Gauss for SGR1806-20.  The observed stability of the
spectra from outburst to outburst argues in favour of an equatorial
emission region (Baring \& Harding 1995; Harding \& Baring 1996).
Splitting can also be quite influential in SGR spectral models in the
absence of cascade formation (e.g.  Thompson and Duncan 1995).

To date, astrophysical models that invoke splitting in the environments
of gamma-ray pulsars (Harding, Baring \& Gonthier 1996, 1997; Chang, Chen
and Ho 1996) and SGRs (e.g. Baring 1995; Baring \& Harding 1995;
Thompson and Duncan 1995; Harding \& Baring 1996; Chang et al. 1996)
use relatively simple approximations to the splitting rates that were
derived by Adler et al. (1970, see also Adler 1971; Bialynicka-Birula
\& Bialynicki-Birula 1970), using effective Lagrangian techniques,
which are valid for low photon energies \teq{\erg m_ec^2} and fields
satisfying \teq{\erg B\ll B_{\rm cr}}.  Restriction to this asymptotic
regime has until recently been necessitated by the total lack of more
general evaluations of splitting rates that are sufficiently amenable
for use in astrophysical models:  formal expressions for the
third-order QED process of \teq{\gamma\to\gamma\gamma} are very
complicated.  This has been a real deficiency, since the \teq{\erg
B\sim B_{\rm cr}} regime can easily be realized in both SGRs and
pulsars.  This paper addresses this limitation by presenting numerical
determinations of splitting rates for arbitrary field strengths and
photon energies below the pair creation threshold of \teq{2m_ec^2}, as
obtained from the general proper-time QED formulation (a technique due
to Schwinger) of splitting presented by Stoneham (1979).

In this paper, our numerical computation of Stoneham's splitting rates
focuses on the polarization mode \teq{\perp\to\parallel\parallel} that
is permitted in the limit of weak dispersion by energy-momentum
kinematic selection rules (see Adler 1971); \teq{\perp\to\perp\perp}
and \teq{\parallel\to\perp\parallel} are the other two modes permitted
by QED in the non-dispersive limit, and results for them are similar to
those presented here (Baring, Harding and Weise, in preparation,
hereafter BHW97).  Here \teq{\perp} and \teq{\parallel} denote the
orientation (perpendicular and parallel to, respectively) of the photon
electric vector relative to the plane containing its momentum and the
field line.  Our computed rates vary continuously from the low energy
limit, and reproduce widely-accepted analytic asymptotic forms at both
low and highly supercritical field strengths.  The numerical
computations presented here therefore appear extremely reliable for use
in astrophysical applications, and concur with the recent numerical
work of Ba\u{\i}er, Mil'shte\u{\i}n, \& Sha\u{\i}sultanov (1996), which
resulted from their alternative Schwinger-type analysis of photon
splitting.  Note, however, that our rates differ dramatically from the
numerical evaluation of the S-matrix determination of the splitting
rate by Mentzel, Berg and Wunner (1994, see also Wunner, Sang and Berg
1995) that is orders of magnitude greater than the earlier proper-time
determinations, and differ by as much as a factor of around 3 from the
recently-updated S-matrix numerics of Wilke and Wunner (1996; see also
the new Erratum to Wunner, Sang and Berg 1995), which still show an
inability to produce accurate results at subcritical field strengths,
the domain applicable to pulsar models.  This discrepancy underlies the
inherent difficulty in computing S-matrix rates for relatively low
fields, and contrasts the reliability of the proper-time computations
presented in this paper and also by Baier et al. (1996).

\section{PHOTON SPLITTING RATES AND SPECTRA}

Magnetic splitting \teq{\split} is a relatively recent prediction of
QED. A decade of controversy followed the earliest calculations before
Bialynicka-Birula \& Bialynicki-Birula (1970), Adler et al. (1970) and
Adler (1971) performed, via an effective Langrangian technique, the
first correct evaluations of its rate, including asymptotic forms in
the limit of photon energies well below pair creation threshold, which
varied as \teq{B^6} when \teq{B\ll B_{\rm cr}}.   Their rate
determinations neglected dispersion in the birefringent, magnetized
vacuum, though the effects of such dispersion on the possibility of
energy and momentum non-conservation and associated polarization
selection rules was discussed in detail by Adler (1971).  The early
controversy was fueled by the inherent difficulties in calculating  the
rates of this third order process by standard QED techniques, and the
physics of photon splitting at high magnetic field strengths is still
the subject of debate.

The most general calculation of photon splitting rates was performed by
Stoneham (1979), who used Schwinger's proper-time method to determine
formal expressions for the splitting rate for \teq{\erg} below pair
creation threshold and arbitrary field strengths.  This analysis fully
included dispersive effects due to magnetized vacuum in formal
expressions for the rates, leaving the rates
for the different polarization modes of splitting as unmanageable
nine-dimensional integrals.   Fortunately, in the limit of zero vacuum
dispersion, Stoneham's formulae specialize to amenable triple integrals
that involve only moderately complicated combinations of elementary
functions (see Adler 1971, Papanyan \& Ritus 1972, and Ba\u{\i}er,
Mil'shte\u{\i}n, \& Sha\u{\i}sultanov 1986 for alternative
presentations, and the recent compact results of Ba\u{\i}er,
Mil'shte\u{\i}n, \& Sha\u{\i}sultanov 1996, hereafter BMS96; Adler and
Schubert 1996).  We have analytically developed these triple integral
expressions (see BHW97) to facilitate our numerical purposes,
eliminating leading-order terms (linear in photon energies) that cancel
exactly, leaving scattering amplitudes that are cubic in photon
energies.  Such cancellations plague numerical evaluations of splitting
rates, and can be a principal source of error.  Hence their elimination
via an algebraic development is highly desirable.  Readers interested
in the details of our analytic reduction of Stoneham's (1979)
expressions can refer to the more extensive presentation in BHW97,
where the full expressions that are computed in this paper are exhibited.

We note that while such analytic developments are expedient and very
useful, it is possible to proceed by numerically subtracting off
cancellations in a well-chosen manner.  This is the approach discussed
in Adler and Schubert [1996; see their Eq.~(17)], who have developed
codes to compute the integrands (i.e. double-integrals) of the triple
integrals composing the scattering amplitude.  Their codes, which are
publicly available on the World Wide Web (the address is listed in
Adler and Schubert 1996), demonstrate numerically the equivalence of
the integrands that emerge from the different formalisms of Adler
(1971), Stoneham (1979), BMS96 and also Adler and Schubert's
presentation.  Unlike the results presented in this paper and the work
of BMS96, Adler and Schubert (1996) do not compute the full triple
integral for the amplitude, a computationally demanding extra step that
is necessary to obtain results of use for astrophysical applications.
The numerical algorithm used to obtain the results of this paper
involves a mixture of Simpson's rule and Gauss-Laguerre quadrature to
obtain accurate differential and total splitting rates.

For quite general photon energies and magnetic field strengths, the
differential attenuation coefficient (i.e. the rate divided by \teq{c})
of the photon splitting mode \teq{\perp\to\parallel\parallel} can be
written in the form (BHW97)
\begin{equation}
T_{\perp\to\parallel\parallel}\; =\; \dover{\fsc^3}{2\pi^2}\,
   \dover{1}{\lambar}\, B^6\, \sin^6\theta\; \erg_1^2\, \erg_2^2\;
  {\cal M}^2_{\perp\to\parallel\parallel}\quad ,
  \label{eq:splitrate}
\end{equation}
for an initial photon of energy \teq{\erg} and produced photons of
energies \teq{\erg_1} and \teq{\erg_2 =\erg -\erg_1}.  Here \teq{{\cal
M}} denotes the scattering amplitude, scaled according to Stoneham's
notation; other polarization modes assume a similar form.  Hereafter,
dimensionless units will be used, with \teq{B} being expressed in units
of \teq{B_{\rm cr}} and photon energies \teq{\erg} (initial) and
\teq{\erg_1} and \teq{\erg -\erg_1} (final) being scaled by
\teq{m_ec^2}.  In equation~(\ref{eq:splitrate}), \teq{\fsc} is the fine
structure constant, \teq{\lambar =3.86\times 10^{-11}}cm is the
electron Compton wavelength over \teq{2\pi}, and \teq{\theta} is the
angle the photon momenta make with the field lines;
equation~(\ref{eq:splitrate}) is strictly valid only in the
zero-dispersion limit, for which the initial and final photon momenta
are collinear.  The triple integral expression we obtained for the
scattering amplitude \teq{\cal M}, somewhat lengthy for presentation
here, has an integrand consisting purely of terms that are cubic
combinations of \teq{\erg -\erg_1} and \teq{\erg_1}, multiplied by an
exponential that is quadratic in these energies, all divided by the
common factor \teq{\erg\erg_1 (\erg -\erg_1)}.

When the incident photon energy is low, namely \teq{\erg\ll 1} (i.e.
511 keV), these energy dependences in the integrand cancel exactly, the
exponential loses its energy dependence and dramatic analytic
simplification is possible.  The scattering amplitude then becomes
dependent only on the field strength:
\begin{equation}
   {\cal M}_{\perp\to\parallel\parallel}\;\approx\; {\cal M}_1(B) 
   \;=\;\dover{1}{B^4}\int^{\infty}_{0} \dover{ds}{s}\, e^{-s/B}\,
   \Biggl\{ \biggl(-\dover{3}{4s}+\dover{s}{6}\biggr)\,\dover{\cosh s}{\sinh s} 
   +\dover{3+2s^2}{12\sinh^2s}+\dover{s\cosh s}{2\sinh^3s}\Biggr\} 
 \label{eq:splitcoeff}
\end{equation}
for arbitrary \teq{B}.  The form of \teq{{\cal M}_1(B)} is just that
given in Adler (1971) and Eq.~41 of Stoneham (1979), and is that used
in the recent astrophysical models of SGRs (e.g. Baring and Harding
1995) and gamma-ray pulsars (e.g. Harding, Baring and Gonthier
1996, 1997).  In the limit of \teq{B\ll 1}, \teq{{\cal M}_1(B)\approx
26/315}, while in the limit of \teq{B\gg 1}, \teq{{\cal M}_1(B) \approx
1/(6B^3)}.  For arbitrary \teq{B} and energies well below pair creation
threshold, a series expansion of the scaled scattering amplitude 
\teq{{\cal M}_{\perp\to\parallel\parallel}} in
terms of photon energy was obtained by BHW97:
\begin{equation}
   {\cal M}_{\perp\to\parallel\parallel}\; =\; {\cal M}_1(B) +
   \biggl[\,\Bigl(\erg_1^2+\erg_2^2\Bigr)\,{\cal M}_{11}(B)+\erg_1\erg_2
   {\cal M}_{12}(B)\,\biggr]\,\sin^2\theta +\dots\quad ,
 \label{eq:loweseries}
\end{equation}
where the produced photons assume dimensionless energies \teq{\erg_1}
and \teq{\erg_2=\erg -\erg_1}, and
\begin{eqnarray}
{\cal M}_{11}(B) &=& 
   -\dover{1}{B^5} \int^{\infty}_{0} ds \, e^{-s/B}\,
   \Biggl\{ \biggl(\dover{21}{64s^2}+\dover{1}{12}\biggr) +
   \biggl(\dover{37}{64s^2}+\dover{47}{96}-\dover{s^2}{40}\biggr)\,
   \dover{1}{\sinh^2s} \nonumber\\
   &&\qquad + \dover{3}{8\sinh^4s} - 
   \biggl(\dover{5}{4s^3}+\dover{s}{30}\biggr)\,\dover{\cosh s}{\sinh s} +
   \biggl(\dover{19}{64s}+\dover{s}{24}\biggr)\,\dover{\cosh s}{\sinh^3s}
      \Biggr\}\quad ,
 \label{eq:M11}
\end{eqnarray}
and
\begin{eqnarray}
{\cal M}_{12}(B) &=&  
   -\dover{1}{B^5} \int^{\infty}_{0} ds \, e^{-s/B}\,
   \Biggl\{ \biggl(\dover{39}{32s^2}-\dover{1}{6}\biggr) +
   \biggl(\dover{35}{32s^2}+\dover{5}{24}-\dover{s^2}{60}\biggr)\,
   \dover{1}{\sinh^2s} \nonumber\\
   &&\qquad + \dover{15}{32\sinh^4s} - 
   \biggl(\dover{13}{8s^3}-\dover{s}{30}\biggr)\,\dover{\cosh s}{\sinh s} + 
   \biggl(\dover{1}{16s}-\dover{5s}{24}\biggr)\,\dover{\cosh s}{\sinh^3s}
      \Biggr\}\quad .
 \label{eq:M12}
\end{eqnarray}
The expansion in Eq.~(\ref{eq:loweseries}) is a result that is quite
useful to pulsar modellers that can be reliably used for
\teq{\erg\lesssim 0.5}.  For \teq{B\lesssim 0.3}, this series reduces
to
\begin{equation}
{\cal M}_{\perp\to\parallel\parallel}\;\approx\; \dover{26}{315}
   +\dover{8B^2}{10395}\,
   \Bigl( 205\erg_1^2+19\erg_1\erg_2+205\erg_2^2 \Bigr)\,\sin^2\theta\quad ,
   \quad \erg_1,\,\erg_2\ll 2/\sin\theta ;\; B\lesssim 0.3\;\; .
   \label{eq:splitseries}
\end{equation}
Recently, Ba\u{\i}er, Mil'shte\u{\i}n, \& Sha\u{\i}sultanov
(1996) have derived the high field limit (i.e. \teq{B\gg 1}) for
splitting at arbitrary energies below pair creation threshold
(\teq{\erg\leq 2}); in the notation used here, their result becomes
(for \teq{\theta =90^\circ})
\begin{equation}
{\cal M}_{\perp\to\parallel\parallel} \;\approx\;
   \dover{1}{B^3\erg\erg_1\erg_2}\,\Biggl\{ 
   \dover{4\erg_1}{\erg_2\sqrt{4-\erg_2^2}}\,\arcsin
      \Bigl(\dover{\erg_2}{2}\Bigr) +
   \dover{4\erg_2}{\erg_1\sqrt{4-\erg_1^2}}\,\arcsin 
      \Bigl(\dover{\erg_1}{2}\Bigr) - \erg\Biggr\}\;\; ,\quad \erg\leq 2\; ,
  \label{eq:splithighB}
\end{equation}
for \teq{\erg_2 =\erg -\erg_1}.  We have reproduced this formula
identically from Stoneham's (1979) formulae and also our algebraic
reduction of his expressions (BHW97).  For photon energies far below
pair creation threshold, equation~(\ref{eq:splithighB}) approaches
\teq{1/(6B^3)}, the high field limit of \teq{{\cal M}_1(B)}, and the
series expansion in energy maps over to the high field limit of
Eq.~(\ref{eq:loweseries}), for which \teq{{\cal M}_{11}\sim -{\cal
M}_{12}\sim 1/(30B^3)}.  The inverse cubic dependence of this result on
the field implies an attenuation coefficient that is virtually
independent of \teq{B} in highly supercritical regimes.

Total splitting attenuation coefficients are obtained by integrating
equation~(\ref{eq:splitrate}) over produced energies \teq{0\leq
\erg_1\leq\erg}.  For \teq{\erg\ll 1} and \teq{B\ll 1}, \teq{{\cal
M}_{\perp\to\parallel\parallel}\to {\cal M}_1(0)=26/315} and  this
integration is trivial, leading to a total attenuation coefficient of
\begin{equation}
   T^{\hbox{\sixrm TOT}}_{\perp\to\parallel\parallel}\; =\;
   \int_0^{\erg} T_{\perp\to\parallel\parallel}\, d\erg_1\;\approx\; 
   T_0\;\equiv\;\dover{\fsc^3}{60\pi^2\lambar}\,\Bigl(\dover{26}{315}\Bigr)^2
   \,\erg^5 B^6\sin^6\theta\;\approx\; 0.116\; \erg^5 B^6\sin^6\theta
   \;\;\hbox{cm}^{-1}\; .
 \label{eq:T0}
\end{equation}
In the
limit of highly supercritical fields (\teq{B\gg 1}), no analytic
evaluation of the total splitting rate [\teq{\propto\erg_1^2\erg_2^2}
times the square of the formula in equation~(\ref{eq:splithighB})] is
apparent, as extensive development leaves some intractable terms.
However, the approximation
\begin{equation}
T^{\hbox{\sixrm TOT}}_{\perp\to\parallel\parallel}\;\approx\;
   \dover{\fsc^3}{32\pi^2}\,\dover{\erg^5}{\lambar}\,\sin^6\theta\; 
   \biggl( \dover{4}{271}-\dover{\eperp}{248}+\dover{9\eperp^2}{988}
   -\dover{7\eperp^3}{1188}+\dover{e^{5\eperp /2}}{3461} \biggr),
   \quad 0\leq\eperp =\erg\sin\theta\leq 2 ;\;\; B\gg 1.
  \label{eq:highBtot}
\end{equation}
is accurate to better than \teq{1.5\%} at all energies below pair
creation threshold, and suffices for numerical purposes.  The
asymptotic formulae represented by equations~(\ref{eq:splitseries})
and~(\ref{eq:splithighB}) and also the approximate result in
equation~(\ref{eq:highBtot}) provide strong checks on the accuracy of
the numerical computations presented here.

\placefigure{fig:ratesamplitude}

The computed attenuation coefficients for the splitting mode
\teq{\perp\to\parallel\parallel} obtained by integrating
equation~(\ref{eq:splitrate}) over produced photon energies
\teq{\erg_1} are displayed in Figure~1a, for different field strengths,
together with the magnetic pair creation rates for the \teq{\perp}
polarization (discussed in Daugherty and Harding 1983; Harding, Baring
and Gonthier 1997).  Two characteristic properties of QED radiation processes
in strong fields are immediately apparent.  First, photon splitting,
like pair creation, is a strongly increasing function of both the
magnetic field strength (when \teq{B\lesssim 1}) and also of the energy
of the incident (absorbed) photon.  Second, at energies near pair
threshold, as a third-order QED process, photon splitting has
attenuation coefficients several orders of magnitude smaller than pair
creation, a first-order process.  Another striking feature of Figure~1a
is how closely the coefficients resemble a \teq{\erg^5} power-law, even
for highly supercritical fields; equation~(\ref{eq:highBtot}) clearly
reveals that the high B asymptotic limit (the dashed curve in Fig.~1a)
has at most modest (i.e. less than a factor of around 2.6) deviations
above the \teq{\erg^5} dependence near pair threshold.  Further,
Adler's (1971) simplest analytic form (\teq{\propto\erg^5B^6}) is quite
accurate below pair creation threshold for \teq{B\lesssim 0.2}.
Deviations from \teq{\erg^5} behaviour are illustrated in Fig.~1b, 
where the ratio \teq{T^{\hbox{\sixrm TOT}}_{\perp\to\parallel\parallel}/T_0}
is plotted as a function of incident photon energy.  For the different
field strengths, quasi-horizontal portions of the curves indicate 
\teq{\erg^5} dependence; clearly from the figure, this simple asymptotic
form is obeyed over a wide range of phase space.  Fig.~1b also demonstrates
the impressive agreement between the numerical results obtained here
and those of BMS96 and Adler (1971).  The intersections of the various
curves with the ordinate axis roughly defines corresponding values
of \teq{(315/26) {\cal M}_1(B)}.

\placefigure{fig:ratesdifferential}

Figure~2 exhibits differential photon production spectra, i.e.
\teq{{\cal M}^2_{ \perp\to\parallel\parallel}}, normalized to unity.  A
remarkable property emerges: except for very near pair threshold, or
for \it highly \rm supercritical fields, the spectra closely
approximate the low energy asymptotic limit of \teq{\erg_1^2 (\erg
-\erg_1)^2/30}.  Magnetic broadening of the spectrum (i.e. with
increasing \teq{B}), a feature of other strong field QED radiation
processes, does however become very significant near pair creation
threshold.  These numerical results are also in accord with the recent
alternative presentation of BMS96.

\section{DISCUSSION}

Motivation for this numerical computation of proper-time results has
been enhanced by a new result on the rates of photon splitting:
Mentzel, Berg \& Wunner (1994, hereafter MBW94) presented an S-matrix
calculation of the polarized splitting rates.  While their formal
development is equivalent (Weise, Baring and Melrose 1997) to an
earlier S-matrix formulation of splitting in Melrose \& Parle
(1983a,b), their presentation of numerical results (see also Wunner,
Sang \& Berg 1995, where the implications of their calculations were
discussed in astrophysical contexts) was in violent conflict with the
splitting results obtained by Adler et al. (1970), Adler (1971),
Bialynicka-Birula \& Bialynicki-Birula (1970), Papanyan and Ritus
(1972), Stoneham (1979) and a number of other more recent works.  The
numerical rates of Mentzel, Berg \& Wunner (1994) for \teq{B=0.1} and
\teq{B=1} are larger by orders of magnitude than our computations (and
also those of BMS96).  Their S-matrix rates exhibit a weak energy
dependence well below pair threshold, and also exceed the pair creation
rate at threshold (\teq{\erg =2}) for \teq{B=0.1}.  Wilke and Wunner
(1996) have very recently retracted the earlier S-matrix rates (see
also the recent Erratum for Wunner, Sang and Berg 1995), confirming a
coding error in the numerics of MBW94.  Their numerical re-evaluation,
a sample of which is depicted in Figure~1a, is much closer to the
proper-time results, but still differs significantly (by as much as a
factor of 3.3 at \teq{\erg =0.1}, \teq{B=1}, and 45\% at \teq{\erg
=1.9}, \teq{B=1} for the illustrated cases) from our determinations,
unless the field is highly supercritical.  The S-matrix and proper-time
techniques should produce equivalent numerical results, and indeed have
done so demonstrably in the case of pair production (see Daugherty \&
Harding 1983;  Tsai \& Erber 1974).  The proper-time analysis, though
non-trivial, is computationally much more amenable than the S-matrix
approach, and has been reproduced in the limit of \teq{B\ll 1} by
numerous authors.  Clearly, the work of Wilke and Wunner (1996), which
does not exhibit any results below about \teq{0.6B_{\rm cr}}, still
demonstrates an inability to produce suitably accurate results at
critical and sub-critical field strengths.  This point is not made
clear in the recent Erratum to Wunner, Sang and Berg (1995).  These
inaccuracies can dramatically impact astrophysical applications to
pulsar models (e.g.  Harding, Baring and Gonthier 1997), where factors
of 3 become very significant.  Such discrepancies underline the
inherent difficulty in computing S-matrix rates for relatively low
fields, where many Landau level quantum numbers for the three
intermediate states must be summed over (e.g. see Weise, Baring and
Melrose 1997), and contrasts the reliability of the proper-time
computations presented in this paper, and by BMS96, that emanate from
relatively simple integrals.

In conclusion, this paper presents the major features of our numerical
computations of proper-time expressions for photon splitting
attenuation coefficients and spectra.  Our numerical results are quite
consistent with several analytic limits obtained by various groups,
including the very recent approach of Adler and Schubert (1996),
comparable to the recent numerical computations of BMS96, and
well-behaved in the light of physical intuition related to strong field
QED processes; they still differ significantly from and represent a
very useful improvement over the S-matrix re-evaluation of Wilke and
Wunner (1996), which becomes problematic at field strengths below
\teq{\sim 4\times 10^{13}}Gauss.  We believe that our proper-time
computations can be reliably and efficiently used in astrophysical
models.

\acknowledgments
We thank Stephen Adler and George Pavlov for reading the paper and
for comments helpful to the improvement of the manuscript.  This work
was supported through Compton Gamma-Ray Observatory Guest Investigator
Phase 5 and NASA Astrophysics Theory Program grants.  MGB thanks the
Institute for Theoretical Physics at the University of California,
Santa Barbara for support (under NSF grant PHY94-07194) during part of
the period in which work for this paper was completed.  We thank Ramin
Sina for making his pair creation code available to us for use in the
preparation of Figure~1a.

\clearpage


\figureout{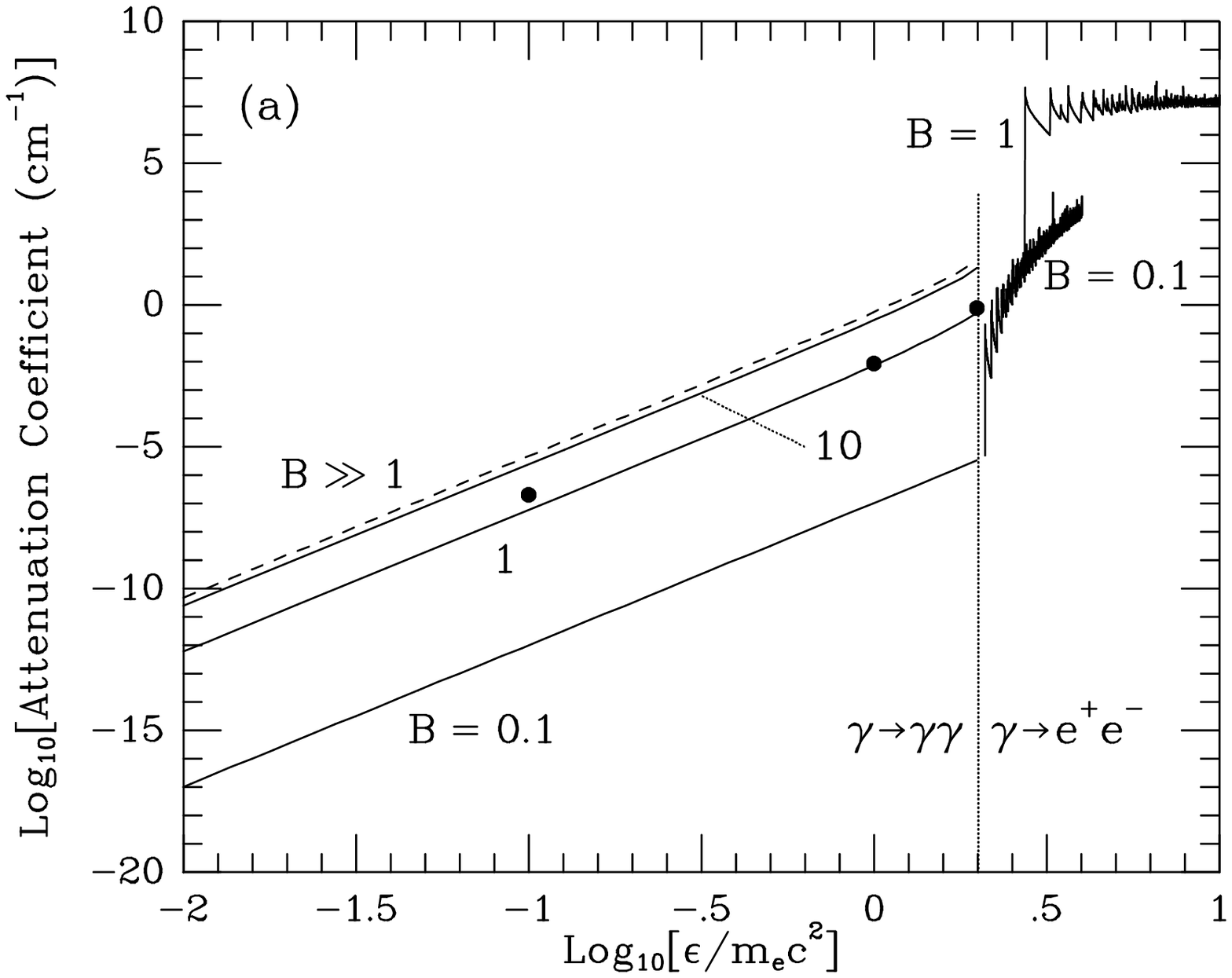}{
   (a) The computed photon splitting attenuation coefficients for
polarization mode \teq{\perp\to\parallel\parallel}, as functions of the
incident photon energy \teq{\erg}, for field strengths \teq{B=0.1,\,
1,\, 10} (solid lines below \teq{2m_ec^2}, in ascending order), and the
asymptotic rate resulting from the insertion of Eq.~(\ref{eq:splithighB})
in Eq.~(\ref{eq:splitrate}), labelled as \teq{B\gg 1} (dashed curve).
Photons are assumed to propagate orthogonally to the field lines
(\teq{\theta =90^\circ}).  For comparison, the pair creation rates
(Daugherty and Harding, 1983) for \teq{\perp} photons at two different
field strengths are depicted above \teq{2m_ec^2}.  
The filled circles denote the S-matrix determination of Wilke and Wunner
(1996) for \teq{B=1}.
    \label{fig:ratesamplitude}
}

\setcounter{figure}{0}
\figureout{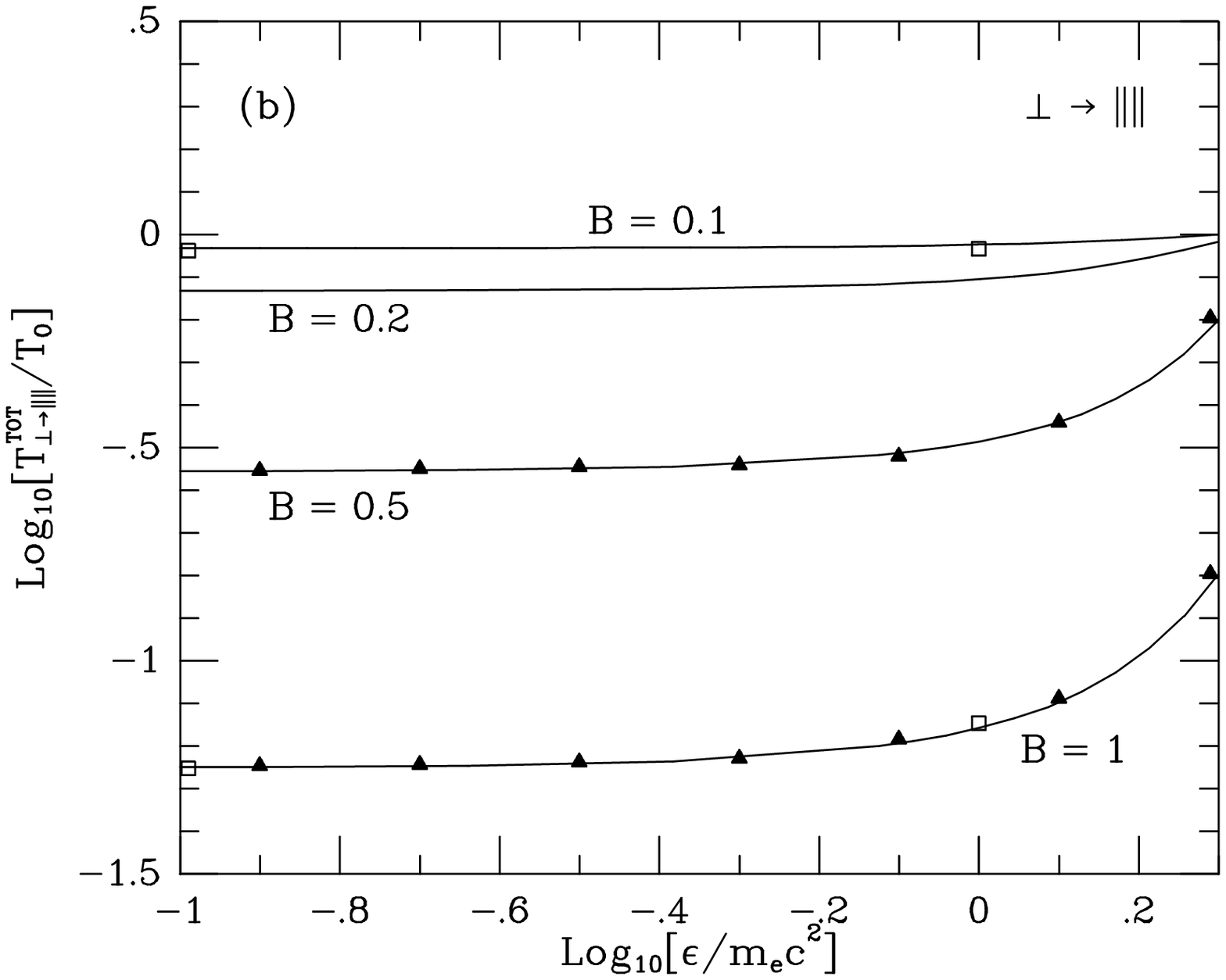}{ 
   (b) The ratio of the total attenuation coefficient 
\teq{T^{\hbox{\sixrm TOT}}_{\perp\to\parallel\parallel}} to the low-energy,
low-magnetic field asymptotic limit \teq{T_0} [see Eq.~(\ref{eq:T0})] as
a function of energy for different field strengths \teq{B}, as labelled.
The open squares represent ratios determined from the computations
of Adler (1971), with the two points near \teq{\erg =0.1} being
the \teq{\erg =0} results from Adler's Fig.~8.  The filled triangles are
derived from the numerical evaluations in Fig.~2 of BMS96.  Impressive
agreement between our results and those of Adler (1971) and BMS96 is
evident.  
} 

\figureout{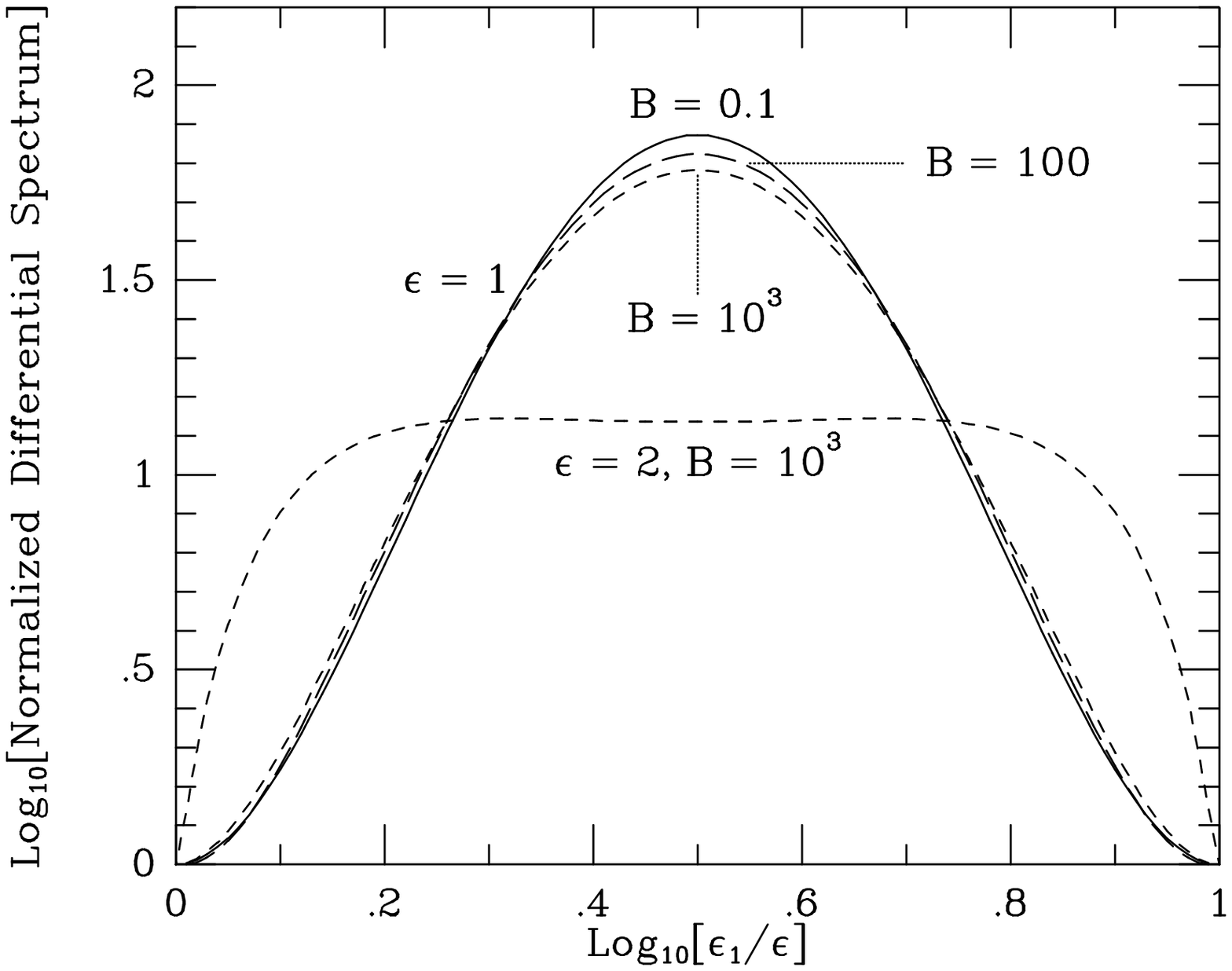}{
    The computed photon splitting differential rates, normalized
to unity and expressed as functions of the energy \teq{\erg_1} of one
of the produced photons, for polarization mode
\teq{\perp\to\parallel\parallel}.  The top three curves are for field
strengths \teq{B=0.1}, \teq{B=100} and \teq{B=10^3} for initial photon
energies \teq{\erg =1} (i.e. 511 keV), with the broad distribution
corresponding to \teq{\erg =2} and \teq{B=10^3}.  The \teq{B=0.1}
curves closely resemble the asymptotic low energy form of \teq{\erg_1^2
(\erg -\erg_1)^2/30}, which can be deduced from
Eq.~(\ref{eq:splitrate}), while the
\teq{B=10^3} distributions closely resemble the shape of the square of
the high field limit in Eq.~(\ref{eq:splithighB}).  
    \label{fig:ratesdifferential}
} 

\end{document}